\documentstyle[12pt]{article}
\begin{document}

\begin{titlepage}
\begin{center}

\bigskip

\bigskip

\vspace{3\baselineskip}

{\Large \bf  New exact cosmological solutions to Einstein's gravity minimally coupled to a Quintessence field}

\bigskip

\bigskip

\bigskip

\bigskip

{\bf Olga Arias, Tame Gonzalez and Israel Quiros}\\
\smallskip

{ \small \it  
Physics Department. Las Villas Central University.\\
Santa Clara 54830. Villa Clara. Cuba}

\bigskip

{\tt  israel@mfc.uclv.edu.cu} 

\bigskip

\bigskip

\bigskip

\bigskip

\vspace*{.5cm}

{\bf Abstract}\\
\end{center}
\noindent

A linear relationship between the Hubble expansion parameter and the time derivative of the scalar field is assumed in order to derive exact analytic cosmological solutions to Einstein's gravity with two fluids: a barotropic perfect fluid of ordinary matter, together with a self-interacting scalar field fluid accounting for the dark energy in the universe. A priori assumptions about the functional form of the self-interaction potential or about the scale factor behavior are not neccessary. These are obtained as outputs of the assumed linear relationship between the Hubble expansion parameter and the time derivative of the scalar field. As a consequence only a class of exponential potentials and their combinations can be treated. The relevance of the solutions found for the description of the cosmic evolution are discussed in some detail. The possibility to have superaccelerated expansion within the context of normal quintessence models is also discussed. 


\bigskip

\bigskip

\end{titlepage}

Dark energy or missing energy is one of the contemporary issues the physics community is more interested in due, mainly, to a relatively recent (revolutionary) discovery that our present universe is in a stage of accelerated expansion\cite{pr}, that was preceded by an early period of decelerated expansion\cite{turner}. This missing component of the material content of the universe is the responsible for the current stage of accelerated expansion and accounts for 2/3 of the total energy content of the universe, determining its destiny\cite{turner}. This is a new form of energy with repulsive gravity and possible implications for quantum theory and supersymmetry breaking\cite{turner}. 

A self-interacting, slowly varying scalar field, most often called quintessence, has been meant to account for the dark energy component. In any successful model of this class, the scalar field energy density should be subdominant at high redshift (in the past) and dominant at low redshift (at present and in the future) in order to agree with observations\cite{turner,peebles}. A variety of self-interaction potentials for the quintessence field has been studied. Among them, the simplest exponential potential (a single exponential) model is unacceptable because it can not produce the transition from subdominant to dominant energy density\cite{peebles}. Other combinations of exponential potentials have been also studied\cite{ratrap,chimento,starobinsky,rs,tmatos}. Combinations of exponentials are interesting alternatives since these arise in more fundamental (particle) contexts: supergravity and superstring\cite{bcn}, where these types of potentials appear after dimensional reduction.

In most cases the occurrence of a self-interaction potential for the scalar field makes difficult to solve analytically the field equations, although some techniques for deriving solutions have been developed. In Ref.\cite{ellis}, for instance, the form of the scale factor is given a priori and, consequently, the self-interaction potential can be found that obeys the field equations. This method has been repeatedly used\cite{uggla,sen} as well as interesting applications of it\cite{saini}. However there are cases when exact solutions can be found once the form of the potential is given\cite{rs,tmatos}. In other cases some suitable relationship between the self-interaction potential and the scalar field kinetic energy is assumed\cite{ygong}. The assumption that the scalar field energy density scales as an exact power of the scale factor has also been used for deriving solutions\cite{jdbarrow}. 

In this paper we explore a linear relationship between the Hubble expansion parameter and the time derivative of the scalar field to derive exact analytic cosmological solutions to Einstein's gravity with two fluids: a barotropic perfect fluid of ordinary matter, together with a self-interacting scalar field fluid accounting for the dark energy in the universe. A similar (but more general) relationship $\dot\phi\propto H$ has been used in Ref. \cite{chimento1} to find general solutions in $N+1$-dimensional theories of gravity with a self-interacting scalar field and also to derive general solutions when we have a self-interacting scalar field with exponential potential plus a free scalar field.\footnotemark\footnotetext{A method where a linear relationship between the field variables and/or their derivatives is assumed has been also used in \cite{acq} to derive 4d Poincare invariant solutions in thick brane contexts.}  The assumed relationship between the Hubble parameter and the time derivative of the scalar field is suggested by an implicit symmetry of the field equations. We will be concerned with flat Friedmann-Robertson-Walker (FRW) cosmologies with the line element given by:

\begin{equation}
\label{line element}
ds^2= -dt^2+a(t)^2 \delta_{ik} dx^i dx^k , 
\end{equation}
where the indexes $i,k = 1,2,3$ and $a(t)$ is the scale factor. 

We point out that it is not neccessary to make any a priori assumptions about the functional form of the self-interaction potential or about the scale factor behavior. These are obtained as outputs of the assumed linear relationship between the Hubble expansion parameter and the time derivative of the scalar field, once one integrates the field equations explicitely. As a consequence only a class of exponential potentials and their combinations can be treated. However, this is not a serious drawback of the method since, as pointed out above, exponential potentials are of prime importance in dark energy contexts accounted for by a quintessence field. The relevance of the solutions found for the description of the cosmic evolution will be discussed in some detail. We use the system of units in which $8\pi G=c=1$.

The field equations are:

\begin{equation}
\label{fec1}
3H^2=\rho_m+\frac{1}{2}\dot{\phi}^2+V, 
\end{equation}

\begin{equation}
\label{fec2}
2\dot H+3H^2=(1-\gamma)\rho_m-\frac{1}{2}\dot\phi^2+V, 
\end{equation} 

\begin{equation}
\label{fec3}
\ddot{\phi}+3H\dot{\phi}=-V', 
\end{equation}
where $\gamma$ is the barotropic index of the fluid of ordinary matter and $H=\frac{\dot a}{a}$ is the Hubble expansion parameter. The dot accounts for derivative in respect to the cosmic time $t$ meanwhile the comma denotes derivative in respect to the scalar field $\phi$. The energy density of the ordinary matter (cold dark matter plus baryons and/or radiation) is related with the scale factor through $\rho_m=\rho_{0,\gamma} a^{-3\gamma}$, where $\rho_{0,\gamma}$ is an integration constant coming from integrating the conservation equation. Let us combine equations (\ref{fec1}) and (\ref{fec2}) to obtain

\begin{equation}
\label{fec4}
\dot H+3H^2= \frac{2-\gamma}{2}\rho_m+2V. 
\end{equation}

An implicit symmetry of the left hand side (LHS) of equations (\ref{fec3}) and (\ref{fec4})is evident under the change $H\rightarrow k\dot\phi$. I. e., if one assumes a linear relationship between the Hubble expansion parameter and the time derivative of the scalar field;

\begin{equation}
\label{relation}
H=k\dot\phi,\Rightarrow a=e^{k\phi}, 
\end{equation}
where $k$ is a constant parameter, the LHS of equations (\ref{fec3}) and (\ref{fec4}) coincide up to the factor $k$. We obtain a differential equation for determining the functional form of the potential $V$:

\begin{equation}
\label{potec}
V'+\frac{1}{k}V=\frac{\gamma-2}{2k}\rho_{0,\gamma}e^{-3k\gamma\phi}. 
\end{equation}

Explicit integration yields the following potential which is a combination of exponentials: $V=\xi_0 e^{-\frac{1}{k}\phi}+\frac{2-\gamma}{6k^2\gamma-2}\rho_{0,\gamma}e^{-3k\gamma\phi}$, where $\xi_0$ is an integration constant. This potential can be given in terms of the scale factor if one considers equation (\ref{relation}):

\begin{equation}
\label{potential}
V=\xi_0 a^{-\frac{1}{k^2}}+\frac{2-\gamma}{6k^2\gamma-2}\rho_{0,\gamma}a^{-3\gamma}. 
\end{equation}

An interesting feature of this potential is that it depends on the type of ordinary fluid which fills the universe. Otherwise, it depends on the barotropic index $\gamma$ of the matter fluid. This fact implies some kind of interaction between the ordinary matter and the quintessence field much like the interacting quintessence studied in Ref.\cite{dpavon}. This potential supports both an early inflation stage and a late time accelerated expansion through a stage of decelerated expansion. By substituting (\ref{potential}) back into Eq. (\ref{fec1}), we can rewrite this last equation in the following form: $(\dot a/a)^2=\frac{2k^2\gamma}{6k^2-2}\rho_{0,\gamma}a^{-3\gamma}+\frac{2k^2\xi_0}{6k^2-1}a^{-1/k^2}$, or, if one introduces the following time variable $dt=a^{1/2k^2}d\tau$,

\begin{equation}
\label{hubble}
(\dot a/a)^2=\frac{2k^2\gamma}{6k^2-2}\rho_{0,\gamma}a^{-3\gamma+1/k^2}+\frac{2k^2\xi_0}{6k^2-1}, 
\end{equation}
where now the dot means derivative in respect to the new time variable $\tau$. This equation can be written in an integral form: $\int\frac{a^{-1}da}{\sqrt{A a^{1/k^2-3\gamma}+B}}=\tau+\tau_0$, or equivalently:

\begin{equation}
\label{integral}
\int\frac{a^{3\gamma/2-1/2k^2-1}da}{\sqrt{A+Ba^{3\gamma+1/k^2}}}=\tau+\tau_0, 
\end{equation}
where $A=\frac{2k^2\gamma}{6k^2\gamma-2}\rho_{0,\gamma}$, $B=\frac{2k^2\xi_0}{6k^2-1}$, and $\tau_0$ is an integration constant. The integral in the LHS of Eq. (\ref{integral}) can be explicitely taken to yield the following expression for the scale factor as function of the time varaible $\tau$;

\begin{equation}
\label{scalef}
a(\tau)=a_0\{\sinh[\mu(\tau+\tau_0)]\}^{2k^2/(3k^2\gamma-1)}, 
\end{equation}
where $a_0=[\frac{6k^2-1}{3k^2\gamma-1}(\frac{\gamma\rho_{0,\gamma}}{2\xi_0})]^{k^2/(3k^2\gamma-1)}$ and $\mu=\sqrt{\frac{(3k^2\gamma-1)^2}{6k^2-1}\frac{\xi_0}{2k^2}}$. The Hubble expansion parameter can then be easily computed:

\begin{equation}
\label{hubble1}
H^2(\tau)=\frac{2k^2}{6k^2-1}\xi_0 a(\tau)^{-1/k^2}\coth^2[\mu(\tau+\tau_0)]. 
\end{equation}

For the purpose of observational testing of the solutions it is useful to look for another magnitudes of astrophysical interest. Among them, the scalar field energy density:

\begin{equation}
\label{senergy}
\rho_\phi=\frac{1}{2}\dot\phi^2+V=3H^2-\rho_m, 
\end{equation}
the scalar field state parameter:

\begin{equation}
\label{omegam}
\omega_\phi=\frac{\frac{1}{2}\dot\phi^2-V}{\frac{1}{2}\dot\phi^2+V}=-1+\frac{1}{3k^2(1-\Omega_m)}, 
\end{equation}
where $\Omega_m=\rho_m/3H^2$ is the density parameter for ordinary matter that is related with the scalar field density paremeter $\Omega_\phi$ through (just another writing for the field equation (\ref{fec1})),

\begin{equation}
\label{omegaphi}
\Omega_\phi=\frac{\rho_\phi}{3H^2}=1-\Omega_m, 
\end{equation}
and last but not the less, the deceleration parameter $q=-(1+\dot H/H^2)$:

\begin{equation}
\label{deceleration}
q=-1+\frac{1}{2k^2}+\frac{3\gamma}{2}\Omega_m. 
\end{equation}

While deriving equations (\ref{senergy})-(\ref{deceleration}) we have used the field equations (\ref{fec1}), (\ref{fec2}), (\ref{fec3}) and their combinations. Assuming the linear relationship (\ref{relation}) between the Hubble parameter and the time derivative of the scalar field means that we are introducing a new free parameter, however, this parameter $k$ can be assumed to be a known function of the other free parameters of the theory ($\tau_0$ and $\xi_0$) or of the barotropic index, etc. Another possibility, perhaps the most promising, is to choose a value for $k$ according to the best fitting of the model to the observations. 

We now investigate the full space of solutions in $k$ parameter without assuming apriori a definite form for it as function of the other parameters $\tau_0$ and $\xi_0$.

A) $k^2<0\Rightarrow\bar k^2=-k^2>0$

This case implies a negative kinetic energy for the scalar field since, due to (\ref{relation}), $\dot\phi^2=-H^2/\bar k^2$. A kinetic energy term with the wrong sign has been advocated in Ref. \cite{caldwell} to produce superaccelerated expansion. In this case the state parameter $\omega_\phi=-\frac{2V+H^2/\bar k^2}{2V-H^2/\bar k^2}<-1$. I. e., in spite of contrary claims \cite{faraoni,dftorres} superaccelerated expansion and, correspondingly, superquintessence can be obtained within the frame of normal quintessence models like this.  As it will be seen later on, from the observational point of view, for large $\bar k^2$ (small negative kinetic energy contribution), superquintessence is undistinguishable from the case with usual accelerated expansion. We want to remark that, although the scalar field kinetic energy is negative, its energy density $\rho_\phi$ could be positive if $2V>H^2/\bar k^2$. Violation of the known energy conditions is obvious since $\omega_\phi=p_\phi/\rho_\phi<-1\Rightarrow p_\phi+\rho_\phi<0$ \cite{dftorres}. 

In terms of the positive parameter $\bar k^2$, the potential (\ref{potential}) takes the following form:

\begin{equation}
\label{potential1}
V=\xi_0 a^{1/\bar k^2}-\frac{2-\gamma}{2(3\bar k^2\gamma+1)}\rho_{0,\gamma}a^{-3\gamma}.
\end{equation}

The scale factor behavior can now be rewritten in the following form:

\begin{equation}
\label{scalef1}
a(\tau)=\bar a_0\{\sinh[\bar\mu(\tau+\tau_0)]\}^{2\bar k^2/(3\bar k^2\gamma+1)},
\end{equation}
where $\bar a_0=[\frac{6\bar k^2+1}{3\bar k^2\gamma+1}(\frac{\gamma\rho_{0,\gamma}}{2\xi_0})]^{\bar k^2/(3\bar k^2\gamma+1)}$ and $\bar\mu=\sqrt{\frac{(3\bar k^2\gamma+1)^2}{6\bar k^2+1}\frac{\xi_0}{2\bar k^2}}$. For the Hubble expansion parameter one has: $H^2(\tau)=\frac{2\bar k^2}{6\bar k^2+1}\xi_0 a(\tau)^{1/\bar k^2}\coth^2[\bar\mu(\tau+\tau_0)]$, meanwhile for the other magnitudes of observational interest:

\begin{equation}
\label{omegam1}
\Omega_m=\frac{3\bar k^2\gamma+1}{3\bar k^2\gamma}\{\cosh[\bar\mu(\tau+\tau_0)]\}^{-2}.
\end{equation}

For the state parameter (equation of state) and the deceleration parameter we have;

\begin{equation}
\label{stateeq}
\omega_\phi=-1-\frac{1}{3\bar k^2(1-\Omega_m)},\;\;\;q=-(1+\frac{1}{2\bar k^2})+\frac{3\gamma}{2}\Omega_m,
\end{equation}
respectively. From this last pair of equations one sees that, for any finite $\bar k^2>0$, the state parameter $\omega_\phi$ is more negative than $-1$, a phenomenon known as superquintessence (see Ref. \cite{caldwell,faraoni,dftorres} and references therein). Besides, an additional non vanishing negative contribution to the deceleration parameter is evident (compare the second equation in (\ref{stateeq}) with (\ref{deceleration})), i. e., superacceleration is obtained. 

We shall notice that some undesirable properties of phantom energy\cite{sahni} are not present or are smoothed out in this case. For instance neither the scale factor of the universe nor the Hubble expansion parameter diverge in a finite (or infinite) amount of cosmic time (pointing into the future). Despite of the fact that, in general, the density of phantom energy $\rho_\phi=\frac{3\bar k^2\gamma\rho_{0,\gamma}\bar a_0^{-3\gamma}}{3\bar k^2\gamma+1}(\sinh^2[\bar\mu(\tau+\tau_0)]-\frac{1}{3\bar k^2\gamma})/(\sinh[\bar\mu(\tau+\tau_0)])^{6\bar k^2\gamma/(3\bar k^2\gamma+1)}$ slowly grows as the universe expands (for large $\bar k^2>>1$, $\rho_\phi\approx\xi_0$), we need an infinite amount of cosmic time to reach an infinite density state.

B) $k^2>0$

We should differentiate two ranges. For $0<k^2<1/3\gamma$ the scale factor evolves periodically:

\begin{equation}
\label{scalef2}
a(\tau)=\tilde a_0 \{\sin[\tilde\omega(\tau+\tau_0)]\}^{2k^2/(3k^2\gamma-1)},
\end{equation}
where $\tilde a_0=[\frac{1-6k^2}{(3k^2\gamma-1)}(\frac{\gamma\rho_{0,\gamma}}{2\xi_0})]^{2k^2/(3k^2\gamma-1)}$. However, in this case, for $0\leq\gamma\leq 2$, $3k^2\gamma-1<0$ so one should have $\xi_0<0$. This means that the self-interaction potential itself is negative, i. e., the scalar field energy density is negative. Besides $q>0$ always and this case is not of observational interest.

For $k^2>1/3\gamma$ the equations (\ref{scalef}), (\ref{hubble1}) are valid and the quantities of observational interest read;

\begin{eqnarray}
\label{observat}
\Omega_m=\frac{3k^2\gamma-1}{3k^2\gamma}\{\cosh[\mu(\tau+\tau_0)]\}^{-2},\nonumber\\
\omega_\phi=-1+\frac{1}{3k^2(1-\Omega_m)},\;\;\;q=-1+1/2k^2+\frac{3\gamma}{2}\Omega_m.
\end{eqnarray}

Now we proceed to observationally test the solution found. We first compute the parameters $\tau_0$, $\xi_0$, and $k$ by the best fit to the observational evidence. These parameters can be fixed once one tries to fit the model with the experimental observations of SN1a (we include the known bounds on $\Omega_{0,m}$, $q_0$, etc.\cite{turner,peebles}), including the recent observation of SN1997ff at redshift $z=1.7$. We are left with a set in parameter space $(\tau_0,\xi_0,k)$ of admissible values fulfilling with the observational bounds. From this set we choose a trio and then we compute the magnitudes of observational interest (equations (\ref{senergy})-(\ref{deceleration})). While doing the computations we used the condition $\rho_{0,m}=n \xi_{0}$. For illustration, in all figures in this paper, we chose the trio ($n=0.5$, $k=10000$, $\xi_0=3$) that agrees with the known bounds on $\Omega_{0,m}$, $\Omega_{0,\phi}$, $q_0$, etc, and also we considered dust matter ($\gamma=1$). For the case $k^2<0 (\bar k^2>0)$ the present values of the state parameter and matter density parameter are $\omega_\phi=-1.00005$ and $\Omega_\phi=0.333341$ respectively, meanwhile, for $k^2>0$, the present values of these parameters are $\omega_\phi=-0.99995$ and $\Omega_\phi=0.333326$.

From fig.1, where the evolution of both $\Omega_m$ and $\Omega_\phi$ $\&$ redshift $z$ is shown, we see that the quintessence field $\phi$ was subdominant in the past as required by nucleosynthesis constraints\cite{fj}, is comparable to the matter density parameter at present $\tau=0 (z=0)$ (according to present observations $\Omega_{0,m}\approx 0.3$ and $\Omega_{0,\phi}\approx 0.7$\cite{turner}) and will be dominant in the future.
 
The dependence of the deceleration parameter on redshif is shown in fig.2 for $k^2<0 (\bar k^2>0)$ and $k^2>0(>1/3\gamma)$. The transition from decelerated expansion into accelerated expansion occurs approximately at $z\sim 0.6$ in both cases, in agreement with claims that the transition should be at $z\sim 0.5$\cite{triess} and with recent observation of the SN1997ff at redshift $z=1.7$, confirming a decelerated phase when the universe was a few seconds old. In fig.3 we show the distance modulus $\delta(z)$ as function of redshift both, computed according to the model (solid line for $k^2<0$ and dashed line for $k^2>1/3\gamma$) and the experimental curve (dots). Relative deviations are of about 0.7$\%$. It is noticeable that both superquintessence and quintessence are undistinguisheable from the observational point of view, and for the values of the parameters considered here. 

The linear relationship explored in this paper allows, also, deriving solutions in Brans-Dicke and non-minimally coupled theories in general and this will be the subject of forthcoming papers.

Summing up. We have derived exact analytic solutions to gravity theory minimally coupled to a self-interacting scalar field by assuming a linear relationship between the Hubble expansion parameter and the time derivative of the scalar field. This relationship is suggested by an implicit symmetry of the field equations. It induces a restriction upon the type of potentials one can deal with: a combination of exponentials that depends on the barotropic index of the matter fluid. We have derived exact analytic solutions and the entire range in $k$ parameter has been explored. It has been shown that,  contrary to common belief, for $k^2<0$ superquintessence behavior is accomodated within this "normal" quintessence model. The solutions found have been observationally tested by using the distance modulus test (SN1a observations where we included the known bounds on $\Omega_{0,m}$, $q_0$, etc.\cite{turner,peebles}) and the SN1997ff observation at $z=1.7$. The main features of present observational cosmology are satisfied within admissible accuracy levels for both $k^2<0$ and $k^2>1/3\gamma$, leaving a margin for superquintessence to effectively take place.


We thank John D. Barrow and Luis P. Chimento for pointing out to us references \cite{jdbarrow} and \cite{chimento1} respectively. We acknowledge the MES of Cuba by financial support of this research.


\newpage

\begin{figure}[b]
\caption{The matter density parameter $\Omega_m$ (circled dotted line) and the scalar field density parameter $\Omega_\phi$ (solid line) as functions of the redshift for $\gamma=1$, $n=0.5$, $k=10000$, and $\xi_0=3$. We have used the condition $\rho_{0,\gamma}=n \xi_0$. The parameters have been chosen such that, at present, $\omega_{0,\phi}=-0.99995$ and $\Omega_{0,m}=0.333326$. It is seen an early stage when the contribution from the quintessence field was subdominant. At present ($z=0$) both contributions from dust and from the scalar field are of the same order ($\Omega_{0,m}=1/3$ while $\Omega_{0,\phi}=2/3$). In the future the quintessence field will be dominant.} 
\end{figure}

\begin{figure}[b]
\caption{Deceleration parameter $q$ as function of the redshift $z$ for $\gamma=1$ and values of the parameters $n=0.5$, $k^2=10000$, and $\xi_0=3$, that agree with the known bounds on $\Omega_{0,m}$, $\Omega_{0,\phi}$, $q_0$, etc. We have used the condition $\rho_{0,\gamma}=n \xi_0$. The curves for $k^2<0$ and $k^2>0(>1/3\gamma)$ are shown toguether (we can not differentiate among them because they coincide). The transition from decelerated expansion into accelerated expansion occurs at $z\approx 0.6$ in agreement with known suggestions and with the observation of the SN1997ff at redshift $z=1.7$} 
\end{figure}

\begin{figure}[b]
\caption{Modulus distance vs redshift for $\gamma=1$ ($n=0.5$, $k^2=10000$, $\xi_0=3$). The solid and the dashed lines represent the results of the theoretical model for $k^2<0$ and $k^2>0(>1/3\gamma)$ respectively. The dots account for the experimental data. A satisfactory agreement is achieved (the relative deviations are of approximately $0.7\%$).} 
\end{figure}


\begin{thebibliography}{99}


\bibitem{pr} S. Perlmutter et al., Astrophys. J. {\bf 517} (1999) 565-586, astro-ph/9812133; A. G. Riess et al., Astron. J. {\bf 116} (1998) 1009-1038, astro-ph/9805201; Astrophys.J. 560 (2001) 49-71, astro-ph/0104455.

\bibitem{turner} M. S. Turner, astro-ph/0202008 (To appear in the Proceedings of 2001: A Spacetime Odyssey (U. Michigan, May 2001, World Scientific)).

\bibitem{peebles} P. J. E. Peebles and Bharat Ratra, astro-ph/0207347.

\bibitem{ratrap} B. Ratra and P. J. E. Peebles, Phys. Rev. D{\bf 37} (1988) 3406.

\bibitem{chimento} L. P. Chimento and A. S. Jakubi, Int. J. Mod. Phys. D{\bf 5} (1996) 71-84, gr-qc/9506015.

\bibitem{starobinsky} A. A. Starobinsky, Grav. Cosmol. {\bf 4} (1998) 88-99, astro-ph/9811360.

\bibitem{rs}  C. Rubano and P. Scudellaro, Gen. Rel. Grav. {\bf 34} (2002) 307-328, astro-ph/0103335.

\bibitem{tmatos} L. A. Urena-Lopez, T. Matos, Phys. Rev. D{\bf 62} (2000) 081302, astro-ph/0003364.

\bibitem{bcn} T. Barreiro, E. J. Copeland and N. J. Nunes, Phys. Rev. D{\bf 61} (2000) 127301, astro-ph/9910214; E. J. Copeland, N. J. Nunes, F. Rosati, Phys. Rev. D{\bf 62} (2000) 123503, hep-ph/0005222.

\bibitem{ellis} G. F. R. Ellis and M. Madsen, Class. Quant. Grav. {\bf 8} (1991) 667.

\bibitem{uggla} C. Uggla, R. T. Jantzen and K. Rosquist, Gen. Rel. Grav. {\bf 25} (1993) 409. 

\bibitem{sen} A. A. Sen, S. Sethi, Phys. Lett. B{\bf 532} (2002) 159-165, gr-qc/0111082.

\bibitem{saini} T. D. Saini, S. Raychaudhury, V. Sahni and A. A. Starobinsky, Phys. Rev. Lett. {\bf 85} (2000) 1162-1165, astro-ph/9910231.

\bibitem{ygong} Y. Gong, Class. Quant. Grav. {\bf 19} (2002) 4537-4542, gr-qc/0203007.

\bibitem{jdbarrow}  C. Rubano and J. D. Barrow, Phys. Rev. D{\bf 64} (2001) 127301, gr-qc/0105037.

\bibitem{chimento1} L. P. Chimento, Class. Quantum Grav. {\bf 15} (1998) 965-974; J. M. Aguirregabiria and L. P. Chimento, Class. Quantum Grav. {\bf 13} (1996) 3197-3209; L. P. Chimento , A. E. Cossarini and N. A. Zuccala; Class. Quantum Grav. {\bf 15} (1998) 57-74; L. P. Chimento, N. A. Zuccala and V. Mendez; Class. Quantum Grav. {\bf 16} (1999) 3749-3763.

\bibitem{acq} O. Arias, R. Cardenas, I. Quiros, Nucl. Phys. B{\bf 643} (2002) 187-200, hep-th/0202130.

\bibitem{dpavon}  W. Zimdahl, D. Pavon and L. P. Chimento, Phys. Lett. B{\bf 521} (2001) 133-138, astro-ph/0105479.

\bibitem{caldwell} R. R. Caldwell, Phys. Lett. B{\bf 545} (2002) 23-29, astro-ph/9908168. 

\bibitem{faraoni} V. Faraoni, Int. J. Mod. Phys. D{\bf 11} (2002) 471-482, astro-ph/0110067.

\bibitem{dftorres} D. F. Torres, Phys. Rev. D{\bf 66} (2002) 043522, astro-ph/0204504.

\bibitem{sahni} V. Sahni and Y. Shtanov, astro-ph/0202346.

\bibitem{ptw} S. Perlmutter, M. S. Turner and M. White, Phys. Rev. Lett. {\bf 83} (1999) 670-673, astro-ph/9901052; M. S. Turner and M. White, Phys. Rev. D{\bf 56} (1997) R4439.

\bibitem{fj} P. G. Ferreira and M. Joyce, Phys. Rev. D{\bf 58} (1998) 023503.

\bibitem{triess} A. Riess, astro-ph/0104455; M. S. Turner and A. Ries, astro-ph/0106051 (Astrophys. J., in press).




\end{thebibliography}
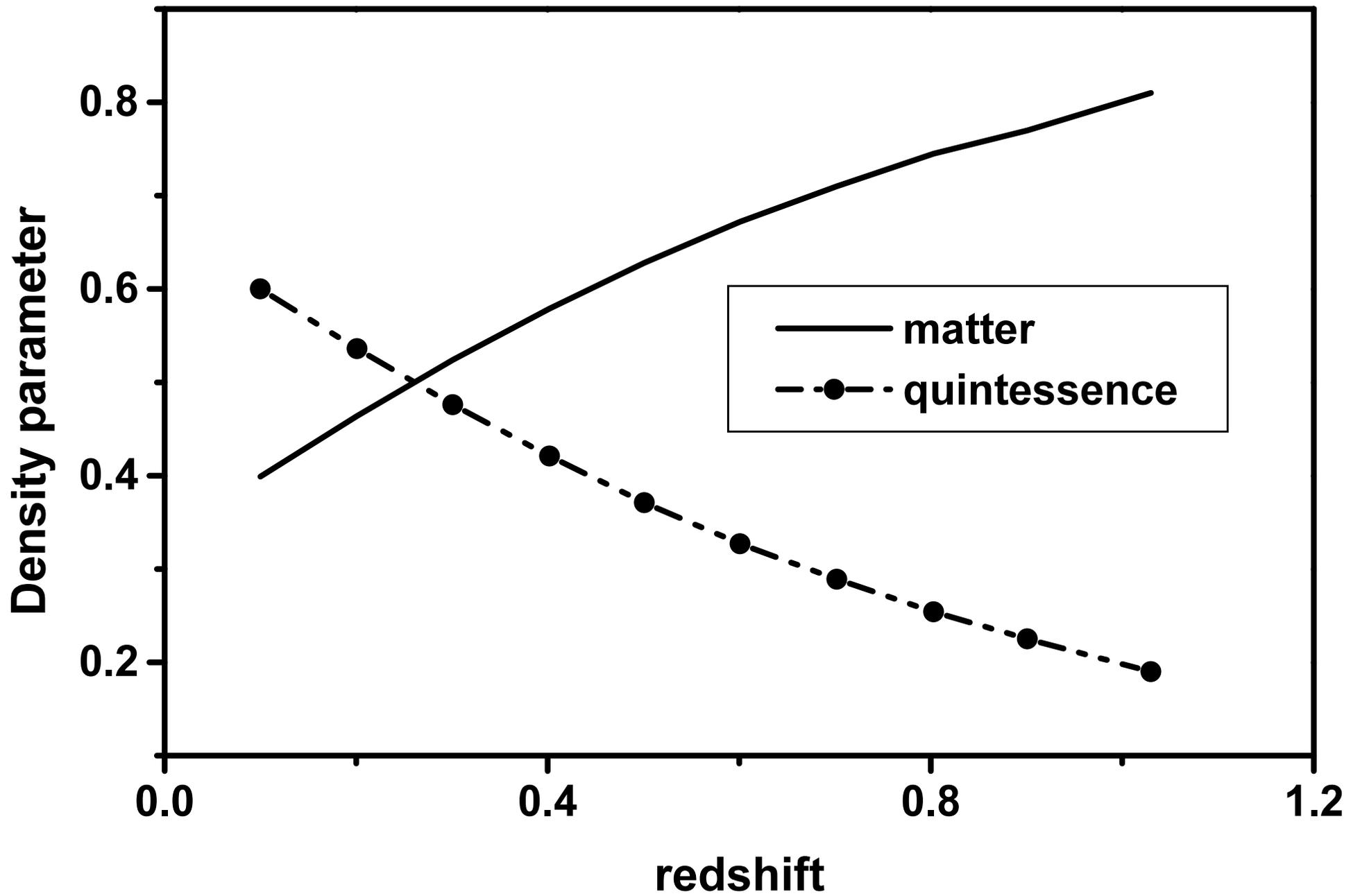
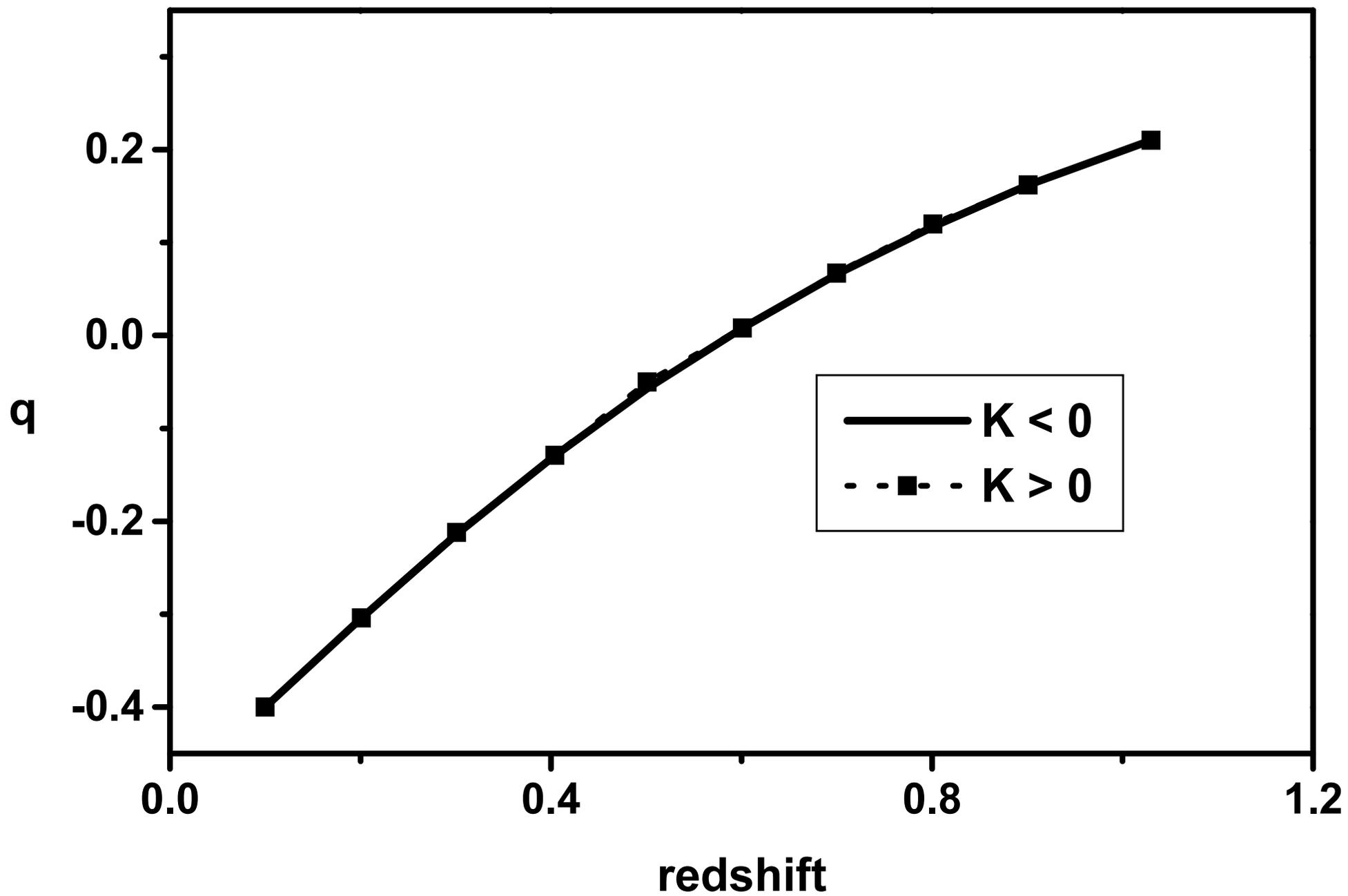
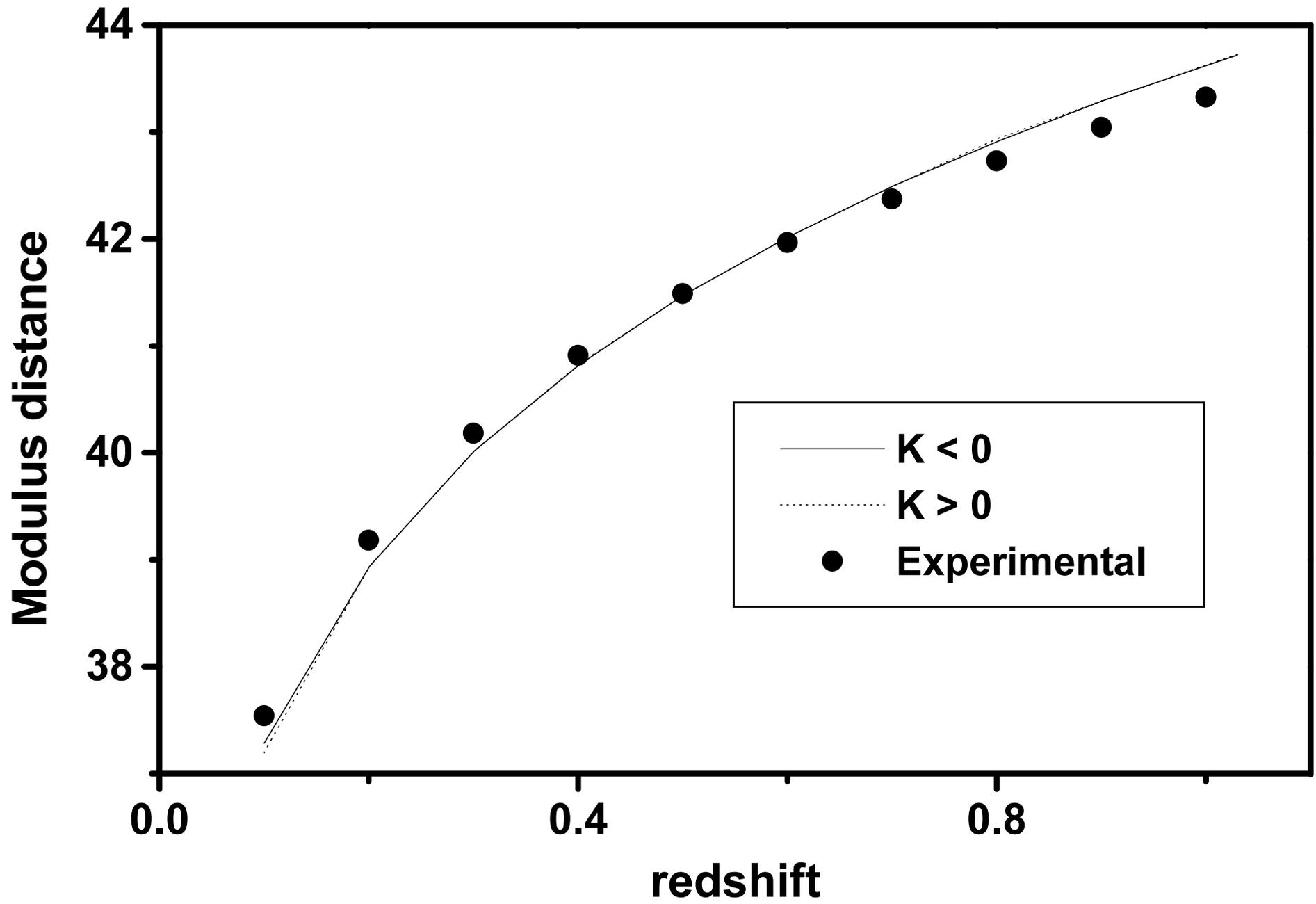
\end{document}